\documentclass[aps,prb,twocolumn,superscriptaddress,showpacs]{revtex4}
\usepackage{graphicx}

\newcommand{\beq}{\begin{equation}}
\newcommand{\eeq}{\end{equation}}

\begin{document}
\title{Domain-Wall Energies and Magnetization of
 the Two-Dimensional Random-Bond Ising Model}

\author{C. Amoruso}
\author{A.K. Hartmann}
\affiliation{Institut f\"ur Theoretische Physik, Universit\"at G\"ottingen,
Tammannstrasse 1, 37077 G\"{o}ttingen, Germany}

\begin{abstract}
We study ground-state properties of the two-dimensional 
random-bond Ising model with couplings having
a concentration $p\in[0,1]$ of antiferromagnetic and $(1-p)$ of
ferromagnetic bonds. We apply
an exact matching algorithm which enables us the
study of systems with linear dimension $L$ up to $700$.
We study the behavior of the domain-wall energies and of the
magnetization. We find   that 
the paramagnet-ferromagnet transition occurs at $p_c \sim 0.103$
compared to the concentration $p_n\sim 0.109$ at the Nishimory point, 
which means
that the phase diagram of the model exhibits a
reentrance. Furthermore, we find  no  indications for an
(intermediate) spin-glass ordering at finite temperature. 
 
{\bf Keywords (PACS-codes)}: Spin glasses and other random models (75.10.Nr), 
Numerical simulation studies (75.40.Mg),
General mathematical systems (02.10.Jf).
\end{abstract}

\maketitle
\section{Introduction} 
Despite more than two decades of intensive research, 
many properties of spin glasses \cite{reviewSG}, especially in finite
dimensions,  are still not well understood. For two-dimensional Ising
spin glasses  it is now widely accepted that no  ordered 
phase for finite temperatures exists
\cite{rieger1996,kawashima1997,stiff2d,houdayer2001,carter2002}.
Furthermore, it seems clear that the behavior can be
described well by a  zero-temperature  droplet scaling approach
\cite{McM,BM,FH,BrayMoore}, but one needs quite large system sizes to
observe \cite{droplet2d} the true behavior.  
One unanswered question is whether an additional phase, usually called
{\em random antiphase}, exists  \cite{Barahona82,MaynardR,Ozeki90,RiegerKaw}
for $T\neq0$ in two dimensions with an asymmetric distribution 
of random bonds. Also it was no clear whether at low temperatures
 the phase boundary of the 
ferromagnetic phase is perpendicular \cite{Nishi_rig,Kitatani}
to the $p$ axis, $p$ denoting the concentration of the
antiferromagnetic bonds. 
The aim of the present paper is to reinvestigate this issues by
studying the domain-wall energy  and magnetization at zero temperature via the
determination of {\it exact} ground states \cite{opt-phys2001}
for large system sizes and huge sample numbers. This allows us to draw
much more reliable conclusions in comparison to past studies,
where only considerable smaller system sizes could be studied.

The organization of the paper is as follows. In the next section, we 
 will expose the model and 
briefly describe the  polynomial matching  algorithm, which allows us
 to treat large system sizes.
Then we will present our results for the
domain-wall energy. In the succeeding section, we explain the
additional methods used to obtain the magnetization and show the
results. Finally, we summarize and draw our conclusions.

\section{The model and the method} 
The model consists of $N=L^2$ spins $S_i = \pm 1$ on a simple 
square lattice with periodic boundary conditions in the $x$ direction
and free boundary conditions in the $y$ direction.
The Hamiltonian  is
\beq
H = -\sum_{\langle i j \rangle}J_{ij}S_i S_j
\eeq
where the sum runs over all pairs of nearest 
neighbors $\langle i j \rangle$ and the $J_{ij}$ are the
quenched random spin-spin couplings. The couplings are set independently
antiferromagnetic ($J_{ij}=-1$) with a probability $p\in[0,1]$ and
ferromagnetic ($J_{ij}=+1$) with probability $(1-p)$.

\begin{figure}[htb] 
\includegraphics[angle=0,width=0.7\columnwidth]{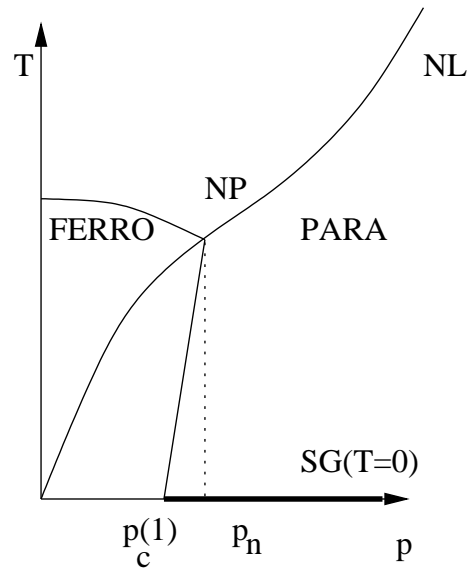}
\caption{The phase diagram of the two-dimensional 
random-bond Ising model, with the
  concentration $p$ of antiferromagnetic bond and the temperature $T$
on the vertical axis. It has been conjectured that the phase boundary
from the $NP$ falls vertically to the $p$-axis.}
\label{pha_dia.fig}      
\end{figure}

The phase diagram of the model as a function of temperature $T$
and the concentration $p$ is shown in 
Fig. \ref{pha_dia.fig}. 
The pure system ($p=0$) has a transition at a Curie temperature
$T_0 = 2[\ln(1+\sqrt{2})]^{-1}$, above which the system is
paramagnetic. When antiferromagnetic bonds are
introduced ($p>0$), the ferromagnetic phase is destroyed at a threshold
concentration $p_c(T)$. A particular curve on the $p-T$ plane is known
as the Nishimori line \cite{nishimori1981} (NL). 
It is defined by the equation $\exp{2\beta} =
\frac{1-p}{p}$. On this line the internal energy is analytic and 
the spin-spin correlation functions  obey the equalities ${\langle
  \sigma_i \sigma_j \rangle}^{2k-1}  = {\langle
  \sigma_i \sigma_j \rangle}^{2k}$, for integer $k$.
It was also proven \cite{LeDoussalHarris}
that a multicritical point  delimiting two critical behaviors on
the ferro-para boundary coincides with the intersection of the $NL$
with the  boundary.
Besides, by studying domain-wall energies of
exact ground states for system's sizes up to  $L=32$,
Kawashima and Rieger \cite{RiegerKaw}
found that the stability of the ferromagnetic and spin-glass 
order cease to exist at a unique concentration for the antiferromagnetic
bonds, so they concluded that there is no intermediate spin-glass
phase.

In this paper we want to compute numerically with high accuracy
the critical  concentration $p_c^{(1)}=p_c(T=0)$ corresponding to the
para-ferro transition at zero temperature, and to compare this result with
the believed value of the Nishimori point $p_n = 0.109(1)$ 
\cite{Honecker-Picco,MerzChalker}.
Furthermore, we want to check with high accuracy, whether there is an
intermediate spin-glass phase at nonzero temperature.

\begin{figure}[htb] 
\includegraphics[angle=0,width=0.95\columnwidth]{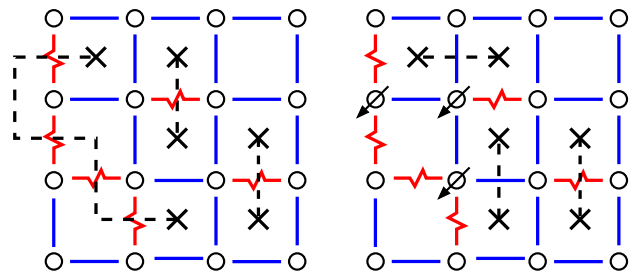}
\caption{2d spin glass with all spins up (left, up spins not
  shown). Straight lines are ferromagnetic, jagged lines
  antiferromagnetic bonds. The dotted lines
  connect frustrated plaquettes (crosses). 
The bonds crossed by the dotted lines
are unsatisfied. In the right part the GS with three spins
pointing down (all other up) is shown, 
corresponding to a minimum number of unsatisfied bonds.}
\label{fig:matching}
\end{figure}

We can reach a much higher precision compared to previous studies, by
applying a matching algorithm. This allows to compute exact ground
states for large system sizes, $N=700^2$ spins in our case.
Let us now explain just the basic idea of the matching algorithm,
for the details, see
Refs. \onlinecite{bieche1980,SG-barahona82b,derigs1991}. 
The method works for spin glasses which
are planar graphs, this is the reason, why we apply periodic boundary
conditions only in one direction. In the left part of
Fig.\ \ref{fig:matching} a small 2d system with open boundary
conditions is shown. All spins are assumed to be ``up'', hence all
antiferromagnetic bonds are not satisfied. If one draws a dotted line
perpendicular to all unsatisfied bonds, one ends up with the
situation shown in the figure: all dotted lines start or end at
frustrated plaquettes and each frustrated plaquette is connected to
exactly one other frustrated plaquette. Each pair of
plaquettes is then said to be {\em matched}. Now, one can consider the
frustrated plaquettes as the vertices and all possible pairs of
connections as the edges of a (dual) graph.
The dotted lines are selected from the edges connecting the vertices and called
a {\em perfect} matching, since {\em all} plaquettes are matched. 
One can assign weights to the edges in the
dual graph, the weights are equal to the sum of the absolute values of the
bonds crossed by the dotted lines. The weight $\Lambda$
of the matching is defined as the
sum of the weights of the edges contained in the matching. As we have seen,
$\Lambda$ measures the broken bonds, hence, the energy of the configuration
is given by $E=-\sum_{\langle i,j\rangle} |J_{ij}|+2\Lambda$. Note
 that this holds for {\em any} configuration of the spins, since a
corresponding matching always exists. Obtaining a ground state means 
minimizing the total weight of the broken bonds (see right panel of
Fig.\ \ref{fig:matching}), so one is looking
for a {\em minimum-weight perfect matching}. This problem is solvable
in polynomial time.

The algorithms for minimum-weight perfect matchings 
\cite{MATCH-cook,MATCH-korte2000} are
among the most complicated algorithms for polynomial problems.
Fortunately the LEDA library offers a very efficient implementation
\cite{PRA-leda1999},  except that it consumes a lot of memory, which
limits in our case 
the size of the systems to about $N=700^2$ on a typical 500 MB
workstation. 

\section{Domain-wall energies and finite-size scaling.} 

We calculate the domain-wall energy $\delta E$ defined by $\delta E \equiv
E_p - E_a$ where $E_p$ and $E_a$ are the ground-state energies with
periodic and the anti-periodic boundary conditions in the
$x$-direction, respectively. We take  an  average over the disorder.  
We are interested in  the exponents $\rho$ and
$\theta_S$ that characterize the system-size dependence of the
mean $\Delta E$ and the width $\sigma(\delta E)$ 
of the  distribution  of the domain-wall energies:
\beq
  \Delta E \propto L^{\rho} \qquad{\rm and}\qquad
  {\sigma(\delta E)} \propto L^{\theta_S}
\eeq
For a general dimensions $d$ of the system, 
a positive  value of $\rho$ indicates the stability of 
a ferromagnetic phase  For $\rho <0$, no
ferromagnetic ordering is present. Then,
in dimension $d$ above the lower critical dimension $d_c$, we have
$\theta_S > 0$ and spin glass-ordering is stable against thermal
fluctuations. On the contrary, when $\theta_S < 0 $, thermal
fluctuations prevent spin-glass ordering \cite{BrayMoore}. The current
believe is that in $d=2$, $\theta_S<0$ holds for all concentrations
$p_c^{(1)}<p<1-p_c^{(1)}$. 

\begin{figure}[htb] 
\includegraphics[angle=0,width=0.88\columnwidth]{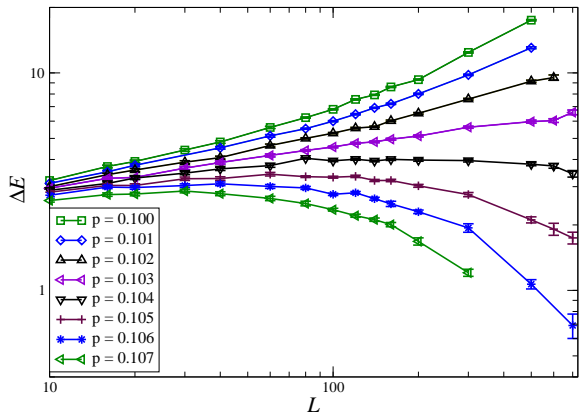}
\caption{The domain-wall energy $\Delta E$ of the bond-random model
  plotted as a function of the system size $L$ for various ferromagnetic-bond
  concentrations $p$.} 
\label{dw_L.fig}
\end{figure}

\begin{figure}[htb] 
\includegraphics[angle=0,width=0.88\columnwidth]{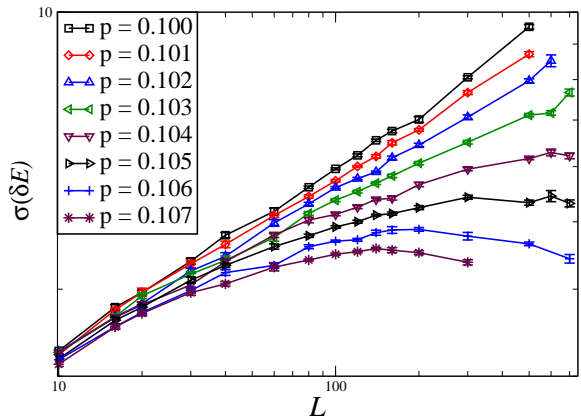}
\caption{The width of the distribution of the domain-wall energy
  $\sigma({\delta E})$ of the bond-random model plotted as a function of the
  system size $L$ for various ferromagnetic-bond  concentration $p$.}
\label{DW_L.fig}
\end{figure}

We have computed  $\sigma({\delta E})$  and $ \Delta E$ for
sizes up to $L=700$ and for values of $p$ ranging from $0.100$
to $0.107$. We performed an disorder average of a number of
realizations ranging from 30000 for the smallest sizes to typically 2000 for the
largest size $L=700$. 
In Fig. \ref{dw_L.fig} and \ref{DW_L.fig}
the mean and the width of the distribution of domain-wall energies
are plotted as a function of the system size. We denote by
$p_c^{(1)}$ and $p_c^{(2)}$ the critical
concentrations of antiferromagnetic bonds at which the asymptotic $L$
dependences of $\Delta E$ and $\sigma({\delta E})$, respectively, change
from increasing to decreasing, i.e., the concentrations where a
ferromagnetic phase and a spin-glass phase, respectively, cease to
exist at finite temperature.  We conclude  from the figures that 
$p_c^{(1)}\sim 0.103 $, while for $p_c^{(2)}$ the ``transition''
is less sharp but the value is between $0.103$ and $0.105$. 
For small sizes, the width even seems to increase for all values of
$p\in[0.1,0.107]$ we have considered.  
For small sizes, at intermediate concentrations 
$p \in [0.1,0.15]$, the mean of the domain-wall energy already
decreases with system size, while the width of the distribution
first increases, which appears as if the system exhibits some kind of
spin-glass phase. This is
probably the reason that in some previous studies
\cite{Barahona82,MaynardR,Ozeki90} the existence of an additional
intermediate phase has been assumed.
We see that we have to consider large system sizes to
observe the true behavior.  

\begin{figure}[htb] 
\includegraphics[angle=0,width=1.0\columnwidth]{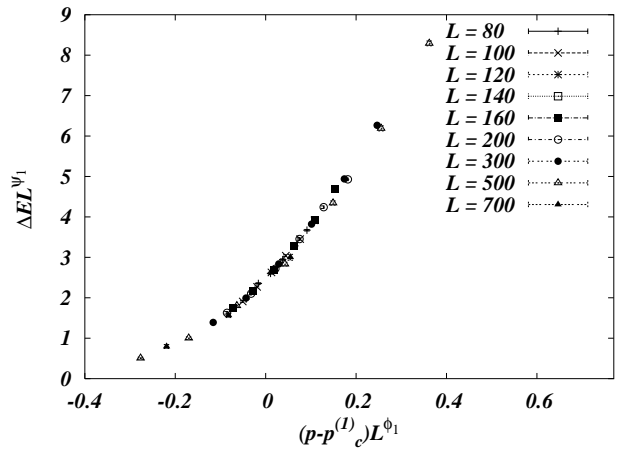}
\caption{The scaling plot of mean domain-wall energy 
$ \Delta E$ versus the concentration of the antiferromagnetic bonds, 
using the values   
$p_c^{(1)} = 0.103$, $\phi_1=0.75$, $\psi_1=-0.12$.}
\label{dw_scal.fig}
\end{figure}
\begin{figure}[htb] 
\includegraphics[angle=0,width=1.0\columnwidth]{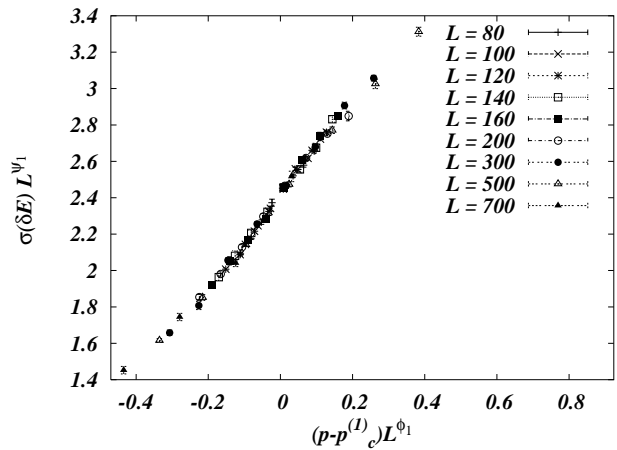}
\caption{The scaling plot of  $\sigma({\delta E})$ versus the
  concentration of the antiferromagnetic bonds,  using the values
  $p_c^{(2)} = 0.104$, $\phi_2 = 0.74$, $\psi_2 = -0.13$.}
\label{DW_scal.fig}      
\end{figure}

\begin{figure}[htb]
\includegraphics[angle=0,width=1.0\columnwidth]{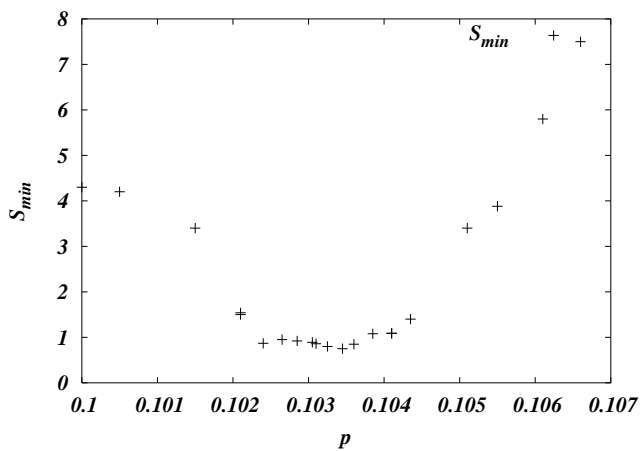}
\caption{Plot of the minimum value of $S(p_c, \psi,\phi)$ for different
  fixed values of $p_c$.}
\label{Smin.fig}
\end{figure}

Another way to compute $p_c^{(1)/(2)}$ is to check the scaling relations
for $\Delta E$ and  $\sigma({\delta E})$  proposed in \cite{RiegerKaw}:
\beq
  \Delta E L^{\psi_1}
  = f_1( (p-p_c^{(1)}) L^{\phi_1} ). \label{eq:FSSI}
\eeq
\beq
  \sigma({\delta E})L^{\psi_2}
  = f_2( (p-p_c^{(2)}) L^{\phi_2} ). \label{eq:FSSII}
\eeq
The  parameters $p_c$, $\phi$ and $\psi$ for both moments 
of domain wall energies have to be
chosen such that a good data collapse for all data is obtained. To
quantify the ``goodness'' of this fit, we used an appropriate cost
function $S(p_c,\phi,\psi)$  introduced in \onlinecite{KawashimaI}  
whose minimum value
should be close to unity when the fit is statistically acceptable.
To minimize $S(p_c,\phi,\psi)$ we used the implementation of the
simplex method  offered by Numerical Recipes library \cite{numrec1995}.
The best fits  give  the estimates
\beq
  p_c^{(1)} = 0.103(1),\quad \phi_1=0.75(5),\quad \psi_1=-0.12(5).
  \label{eq:EstimateBondI}
\eeq
with $S = 0.75$ and
\beq
  p_c^{(2)} = 0.104(2),\quad \phi_2 = 0.74(5),\quad \psi_2 = -0.13(5)
  \label{eq:EstimateBondII}
\eeq
with $S = 0.65$. The resulting scaling plots are shown in
Figs.\ \ref{dw_scal.fig} and \ref{DW_scal.fig}. 
We have estimated the error bars given above in the following way. For each
parameter, we fix it to different values and perform the minimization
over the remaining two parameters.  
In Fig. \ref{Smin.fig} we show as example 
a plot of this partly minimized value of $S(p_c,\psi,\phi)$
as a function  of $p_c$. Our error bars are the ranges of values where
$S(p_c,\psi,\phi)$ increases to twice of its minimum value. 

Within the statistical errors the critical
parameters for both moment of $\delta E$ agree: this
strongly suggest the absence  of a spin-glass phase.
Therefore there is  a  discrepancy between the critical concentration $p$ 
evaluated at the $NP$ and at zero temperature, in disagreement
with the conjecture $p_n=p_c^{(1))}$ by Nishimori \cite{Nishi_rig} and later 
by Kitatani \cite{Kitatani}.
Later, Le Doussal and
Harris {LeDoussalHarris} have shown
that the tangent  to the phase boundary at the $NP$ is vertical.
But this result does not exclude the possibility of a reentrance
in the phase diagram as shown in Fig. \ref{pha_dia.fig}.

\section{Study of the magnetization.}

We furthermore study the para-ferro transition by evaluating the
magnetization and using the Binder cumulant
crossing method \cite{binder81,bhatt85}. The Binder cumulant is given by 
\beq
B(p,L)  = \frac{3}{2}\left(1-\frac{\langle m \rangle^4}{{\langle m^2
      \rangle}^2}\right)\,\,, 
\eeq
where $m = \frac{1}{N}\sum{S_i}$ is the magnetization and $\langle
\ldots \rangle$ denotes the average over the disorder.
For second order phase transitions, the curves for different sizes 
intersect at one point, the
critical concentration $p_c^{(1)}$. This is a consequence of
finite-size scaling.  

The problem one has to face when studying system with discrete interactions,
is the exponentially large number of states all giving the same
energy. Hence there is no unique ground-state magnetization.
For a given set of bonds we here determine {\em one} exact ground state using
an efficient polynomial time  ``matching'' algorithm, but 
we are not able to enumerate all the ground states \cite{LandCopp}.

In order to check if ``typical'' configurations with respect to the
magnetization are found, we first performed
a zero-temperature Monte Carlo (MC) simulation which consists in flipping 
all spins with zero local field, starting with the ground-state
configuration. This allows to explore all states within a
single-spin-flip cluster of ground-state configurations (see below).
In Fig \ref{mcs.fig} a typical
evolution of the magnetization, here for $L=300, p=0.1$,
averaged over $100$ samples, is shown as a function of the number of Monte
Carlo steps. We observe that 
after few hundred MC steps the simulation is equilibrated. Additionally,
we see that the value of the magnetization found
by the ground-state algorithm, i.e. for zero MC steps, 
is slightly outside the scattering of the datapoints for large Monte
Carlo steps.
But compared to the final statistical error bars 
(see below), this difference is negligible. Hence we conclude that the
ground state obtained by the matching algorithm exhibits a typical
magnetization of the cluster, in which the ground state is located.
\begin{figure}[htb] 
\includegraphics[angle=0,width=1.0\columnwidth]{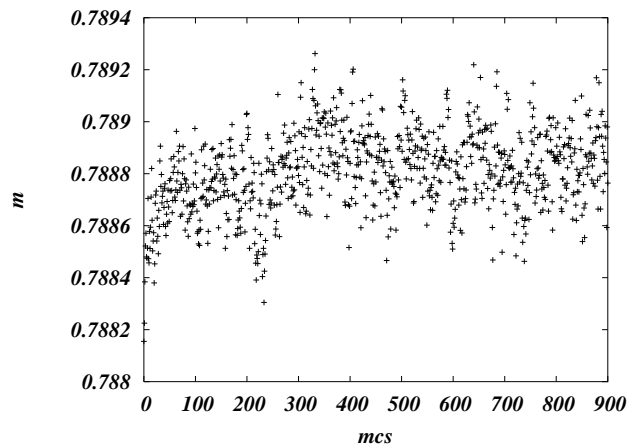}
\caption{Magnetization averaged over 100 samples ($L=300$, $p=0.1$)
as a function of the number of MC
  steps for a zero-temperature single-spin-flip dynamics.}
\label{mcs.fig}
\end{figure}

Anyway, the set of ground states usually is divided into {\em several} 
clusters \cite{sg-clusters}: 
different ground states belong to the same cluster if they are related by 
a sequence of single spin flips, each leaves the energy the
same. Ground states in  different cluster can only be reached from each
other by making cooperative flips of multiple spins or when using
single-spin flips via increasing the energy. This means that
with our single-spin-flip MC algorithm at $T=0$ 
the system always stays in the same cluster.
In principle one can enumerate all ground states
\cite{LandCopp}. Since the ground-state degeneracy grows exponentially
fast with the system size, only small systems can be treated like this.

\begin{figure}[htb] 
\includegraphics[angle=0,width=1.0\columnwidth]{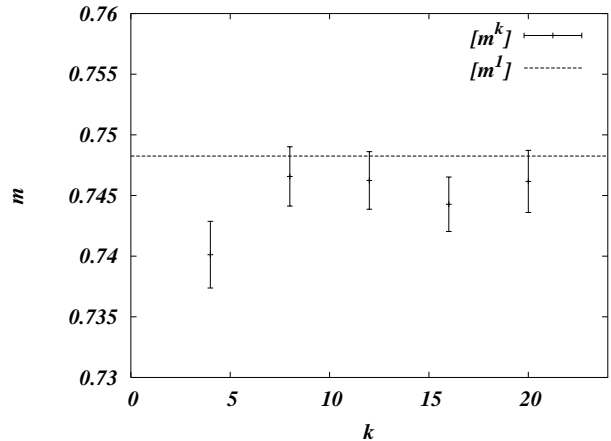}
\caption{Magnetization obtained after averaging over $k$ independent
  ground states as a function of $k$. The horizontal lines indicate
  the magnetization of the first ground state $k=1$. The data is for
  $L=100$, $p=0.103$ and averaged over 1000 realizations.}
\label{eps_coupling.fig}
\end{figure}

Thus, we have applied an  alternative method to find different ground
states, the 
so called {\em $\epsilon$-coupling method}. It allows to obtain ground states
from different clusters, but no exhaustive enumeration is necessary.
The basic idea is to first add a perturbation to
the system which tends to increase the energy if two neighbouring 
spins are in the same relative orientation like in
the ground state 
and then to recalculate the ground state.
 Let $S_i^{(1)}$ be the ground-state
spin configuration. We then perturb the couplings $J_{ij}$ by an amount
proportional to $S_i^{(1)} S_j^{(1)}$ in order to repel the system from the 
ground state. This perturbation, which depends upon a positive
parameter $\epsilon$, is defined by 
\beq
J_{ij}\rightarrow J^{}_{ij} + \Delta J_{ij} 
\eeq
where
\beq
\Delta J_{ij} = -\frac{\epsilon}{N_b}S_i^{(1)}S_j^{(1)}\,
\eeq
where $N_b$ is the number of bonds in the system.
We then recompute the ground state and check that the new configuration
is still  a ground state of the unperturbed Hamiltonian. This is our
second ground state ${S_i^{(2)}}$. 
 The next step, to obtain a third ground state, consists in adding  a 
perturbation in order to repel the system from both ground states
obtained so far. This process can be iterated. 
For a number $n$ of steps of this process, we have:
\beq
\Delta J_{ij} = 
-\alpha \sum_{k=1}^{n}\frac{\epsilon}{N_b}S^{(k)}_{i}S^{(k)}_{j} ,
\eeq
where $\alpha$ is a scaling factor that we choose equal to $1$. 
In this way we hope to find configurations  belonging to
different clusters, even if this procedure is not completely under control
since it is a biased random sampling in the space of configurations.
To test the behavior of our method, we have calculated
for each step $k$ of the $\epsilon$-coupling approach the
magnetization $m^{k}$ of the $k$th ground state. 
Fig. \ref{eps_coupling.fig} shows the resulting behavior of  
$[{m^{k}}]$, $[\ldots]$ denoting the
average over the first $k-1$ iterations of the $\epsilon$-coupling method 
and over 1000 configurations the disorder ($L = 300$, $p=0.1$).
We observe that, within the statistical error bars, the magnetization for
the first ground state, indicated by the horizontal lines,
 agrees well with the value obtained after
averaging over several different degenerate ground-state configurations.
Note that for small $k$ the difference is larger. This is due to the
fact that the $\epsilon$-coupling method repels each configuration
from the previously obtained ground states. Hence, for small $k$ it
will move strongly away from the first ground state. With increasing
number $k$ of steps, the different ground states will be scattered
around in configuration space. If the $k=1$ ground state is typical
with respect to the magnetization, then the average will converge to
the initial value again, which seems to be the case here. Note that we
have also checked for the different iterations of the 
$\epsilon$-coupling method that performing an additional $T=0$ MC simulation
changes the obtained magnetization values again only slightly.
Since the $\epsilon$-coupling method strongly slows the simulations and the
differences are negligible within statistical error bars, 
we have restricted the simulations to the
immediately obtained ground states ($k=1$), i.e. without applying the
$\epsilon$-coupling method.

\begin{figure}[htb] 
\includegraphics[angle=0,width=0.88\columnwidth]{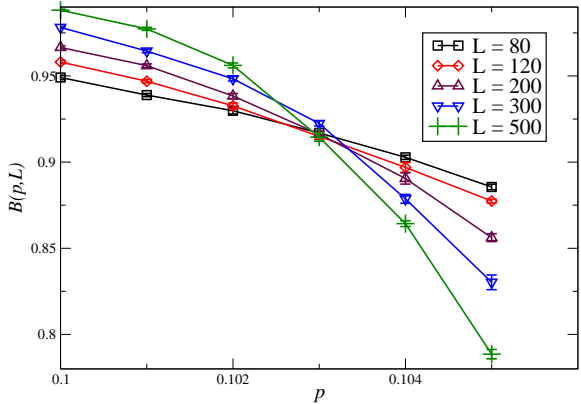}
\caption{Binder cumulant $B(p,L)  = \frac{3}{2}(1-{\langle m\rangle
    ^4}/{\langle m^2\rangle^2})$ as a function of the concentration of
  antiferromagnetic bonds $p$.}
\label{binder.fig}
\end{figure}

In Fig. \ref{binder.fig} the Binder cumulant  is shown as a function
of the concentration of $p$ for different sizes
$L\le 500$, where we have obtained data for all concentrations 
$p=0.1,$ $0.101,$ $\ldots,$ $0.107$. 
It intersects close $p_c^{1}  \sim 0.103$ expect for the largest
size, where the stistics is not so good. Hence, we again conclude
$p_c^{1}= 0.103(1)$ which
agrees well with what we have obtained above 
by studying domain wall energies.
This also indicates that the matching algorithm indeed finds ground states
which are typical with respect to the magnetization.

\begin{figure}[htb] 
\includegraphics[angle=0,width=1.0\columnwidth]{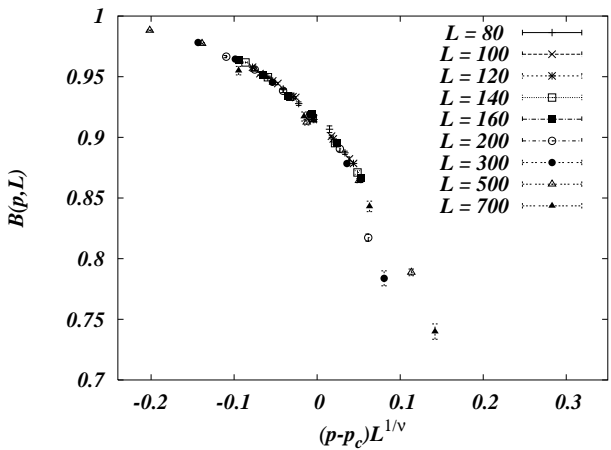}
\caption{Scaling plot of the Binder cumulant
  $B(p,L)$ as a function of $(p-p_c)L^{{1}/{\nu}}$ with $p_c  =
  0.103$ and $\nu = 1.55$.}
\label{binder_fss.fig}
\end{figure}

Finally, we present  further ways to check the previous conclusions. 
We have performed another treatment of the data by trying
to collapse  all curves in Fig. \ref{binder.fig} in a single 
one. This means we want to find the parameters which satisfy the 
finite-size scaling relation for the Binder cumulant:
\beq
B(p,L) = \tilde{B}(L^{{1}/{\nu}}(p-p_c))\,\,
\eeq 
 where $\nu$ is the critical exponent of the correlation length.
We vary $p_c$ and $\nu$ in order to minimize the functional
$S(p_c,\nu)$. We find 
its minimum value for $p_c = 0.103$ and $\nu=1.55(1)$
($S\sim 2.2$), the resulting data collapse is illustrated in
Fig. \ref{binder_fss.fig}. The value for $\nu$
is consistent with the value $\nu=1.50(3)$
at the $NP$ reported in Ref. \onlinecite{MerzChalker},
but differs from the value $\nu=1.33(3)$ found previously
\cite{Honecker-Picco}. Since in the work of Merz and Chalker
 much larger system sizes
are treated compared to the work of Honecker et. al., the value $\nu=1.50(3)$
appears more reliable. Hence, our result indicates  that the 
transition at $NP$ and at $T=0$ are in the same universality class.

\begin{figure}[htb] 
\includegraphics[angle=0,width=1.0\columnwidth]{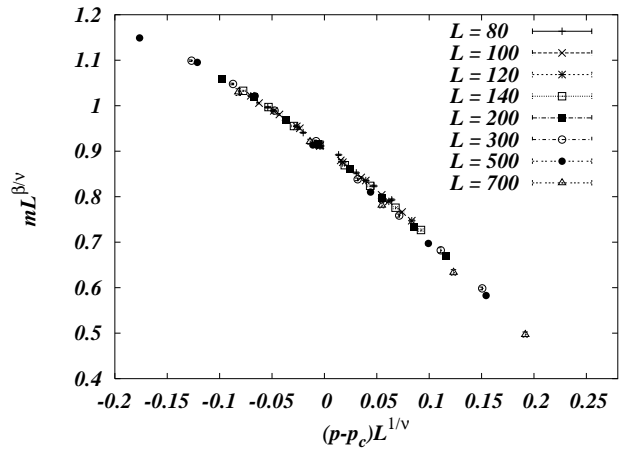}
\caption{Scaling plot of the rescaled magnetization
  $mL^{{\beta}/{\nu}}$ 
as a function of $(p-p_c)L^{{1}/{\nu}}$ with $p_c  =
  0.103$, $\nu = 1.55$, $\beta=0.9$.}
\label{m_scal.fig}
\end{figure}

Finally, we study the finite-size scaling behavior of the magnetization.
The prediction  for the magnetization is
\beq
m(p)  = L^{-{\beta}/{\nu}}\tilde{m}((p-p_c)L^{{1}/{\nu}})\,
\eeq
where $\beta$ is the critical exponent of the magnetization.
Using $p_c = 0.1032$, $\nu =1.55$, a good data collapse
is obtained with $\beta = 0.9(1)$, see Fig. \ref{m_scal.fig}.

\section{Conclusions}
To summarize, we have performed a systematic calculation of the 
domain-wall energy at zero temperature for the two-dimensional
random-bond Ising model for different concentrations $p$ of the
antiferromagnetic bonds. By using a matching algorithm, we could study
systems which are much larger than in previous studies.

We find that both  ferromagnetic and 
spin-glass phases cease to exist at the same concentration $p_c=
0.103(1)$. This means, we do not find any sign for an intermediate
``random antiphase''. Furthermore,
the values of $p_c$ at $T=0$ and at the Nishimory point, found in the
most reliable studies so far \cite{Honecker-Picco,MerzChalker}, are different,
indicating a reentrance of the paramagnetic phase.

Large system sizes $L\ge 500$ are needed to observe the true thermodynamic
behavior with good accuracy. 
Slightly above $p_c$, the width of the distribution of
domain-wall energies increases for small sizes, while it decreases for
larger sizes. Note that in principle we cannot exclude that a
similar turnover happens for smaller concentrations, e.g. $p=0.103$,
at even larger system sizes $L>700$, which are out of reach of our
algorithm. Nevertheless, this would mean that the real $p_c$ is even
smaller, hence the discrepancy between $p_c$ and $p_n$ would
increase and the reentrance became stronger.

\section{Acknowledgments}
The authors have obtained financial support from the
{\em VolkswagenStiftung} (Germany) within the program
``Nachwuchsgruppen an Universit\"aten'' and from the European Community
via the Human Potential Program under contract number
HPRN-CT-2002-00307 (DYGLAGEMEM), via the High-Level Scientific
Conferences (HLSC) program, and
via the Complex Systems Network of Excellence ``Exystence''.
The simulations were performed at the Paderborn Center
for Parallel Computing in Germany and on a workstation
cluster at the Institut f\"ur Theoretische Physik, Universit\"at
G\"ottingen,
Germany.

\end{document}